\documentclass{article}
\usepackage{cite}
\usepackage{tabularx} 
\usepackage{graphicx}
\usepackage{subcaption}

\title{Usability of Token-based and Remote Electronic Signatures: A User Experience Study}
\author{Ömer Ege, Mustafa Çağal, Kemal Bıçakcı}
\date{April 2025}
\usepackage{hyperref}

\begin{document}

\maketitle

\begin{abstract}
As electronic signatures (e-signatures) become increasingly integral to secure digital transactions, understanding their usability and security perception from an end-user perspective has become crucial. This study empirically evaluates and compares two major e-signature systems—token-based and remote signatures—through a controlled user experience study with 20 participants. Participants completed tasks involving acquisition, installation, and document signing using both methods, followed by structured surveys and qualitative feedback. Statistical analyses revealed that remote e-signatures were perceived as significantly more usable than token-based ones (\(p < 0.001\)), due to their minimal setup and platform-independent accessibility. In contrast, token-based signatures were rated as significantly more secure (\(p < 0.01\)), highlighting users’ trust in hardware-based protection. Although more participants preferred remote e-signatures for document signing, the preference did not reach statistical significance (\(p = 0.058\)), indicating a trend toward favoring convenience in real-world scenarios. These findings underline the fundamental trade-off between usability and perceived security in digital signing systems. By bridging the gap between theoretical frameworks and real user experience, this study contributes valuable insights to the design and policy-making of qualified electronic signature solutions.
\end{abstract}


\section{Introduction}

The rapid advancement of digital technologies has transformed many sectors, and electronic signatures (e-signatures) have become a crucial element in digital identity verification, document authentication, and online transactions. E-signatures allow individuals and organizations to authenticate documents remotely, reducing the need for physical paperwork and enabling more efficient and cost-effective business practices. Qualified electronic signatures (QES) are necessary for the safe authentication of individuals in a number of transactional e-government services \cite{theuermann2019}. Although the adoption of electronic signatures has increased significantly in various industries such as finance, e-commerce, and e-government, the usability of these systems remains a significant concern\cite{marshall2025relevance},\cite{Lax01072015}. Despite their security and legal validity, many users struggle with the complexity of the processes involved in acquiring, installing, and using electronic signature systems\cite{cagal}.

Qualified Electronic Signatures (QES) are increasingly critical for secure digital transactions; however, previous studies have largely focused on system design and legal frameworks rather than end-user experiences. Among these, the work of Çağal and Bıçakcı~\cite{cagal} provided a significant contribution by systematizing QES use cases and identifying usability challenges through cognitive walkthroughs. Their research primarily emphasized the conceptual categorization of different QES implementations and highlighted potential usability barriers based on expert evaluations.

Building upon this foundation, the present study aims to empirically assess and compare user experiences with token-based and remote electronic signature systems. By conducting a controlled user study involving real participants, this research moves beyond theoretical analyses and provides concrete evidence regarding the usability and security perceptions associated with these two signature methods. Thus, it addresses a critical research gap by transitioning from systematized design paradigms to user-centered experimental validation.

\section{E-Signature Methods and Their Implementation}

Electronic signatures (e-signatures) ensure authentication and verification of digital documents and data. According to the European Union's regulation \textit{ eIDAS} (Electronic Identification, Authentication, and Trust Services), e-signatures are classified as follows \cite{eIDAS}:

\begin{itemize}
    \item \textbf{Simple Electronic Signature (SES)}: Basic verification methods associated with digital documents (e.g., scanned signatures, email confirmations).
    \item \textbf{Advanced Electronic Signature (AES)}: A signature uniquely linked to the signatory, ensuring identity verification and document integrity.
    \item \textbf{Qualified Electronic Signature (QES)}: The highest legally recognized e-signature issued by a trusted service provider (TSP) based on qualified certificates.
\end{itemize}

Token-based electronic signatures typically fall under the category of qualified electronic signatures (QES). Remote electronic signatures can qualify as QES if they are created by a Qualified Trust Service Provider (QTSP) using secure signature creation devices (QSCD) or equivalent secure environments. According to the eIDAS Regulation (EU No 910/2014), a remote signature meets the QES requirements if it ensures the same level of security as a local QSCD \cite{eIDAS}.
However, as of 2025, remote qualified electronic Signatures are not yet legally recognized in Türkiye. Local regulations require the use of physical qualified signature creation devices (QSCD) for QES compliance \cite{turkiyeqes}.

\subsection{Remote Electronic Signatures}
\subsubsection{Definition and Technological Infrastructure}
Remote electronic signatures allow users to authenticate and sign documents online without requiring physical devices. Identity verification is performed through methods such as

\begin{itemize}
    \item Video identification,
    \item Mobile authentication (e.g., OTP-based verification),
    \item e-ID (electronic identity card) verification.
\end{itemize}

The signature creation data are securely stored in cloud-based infrastructures managed by trusted service providers, following standards such as ETSI TS 119 432~\cite{etsi2019}.

\textbf{Advantages:}
\begin{itemize}
    \item Device-independent: Accessible via computers, tablets, or smartphones.
    \item No installation required: No software or driver installation is required.
    \item High accessibility: Can be used from anywhere.
\end{itemize}

\textbf{Disadvantages:}
\begin{itemize}
    \item Security concerns: Vulnerable to phishing attacks and credential theft. Initiatives like the Cloud Signature Consortium have proposed standardized frameworks to enhance the security and interoperability of cloud-based electronic signature services~\cite{cloudsignature2016}.

    \item Limited legal recognition: Not legally accepted in all jurisdictions (e.g., not yet recognized in Türkiye).
\end{itemize}

\subsubsection{Acquisition Process}
\begin{enumerate}
    \item \textbf{Application}: Users apply online through a trusted service provider.
    \item \textbf{Identity Verification}: Users undergo verification via video call, biometric authentication, or bank eID integration.
    \item \textbf{Certificate Issuance}: A remote signature certificate is generated and securely stored in the provider cloud.
    \item \textbf{Activation}: Users receive an OTP or mobile notification to activate their signature.
\end{enumerate}

\subsubsection{Usage Process}
\begin{enumerate}
    \item \textbf{Signing}: The user uploads the document to the platform and authorizes the signature via OTP or biometric confirmation.
    \item \textbf{Verification}: Signed documents are validated through TSPs or eIDAS-compliant verification systems.
    \item \textbf{Management}: Users access and manage signed documents via the service provider’s online portal.
\end{enumerate}

\subsection{Token-Based Electronic Signatures}
\subsubsection{Definition and Technological Infrastructure}
Token-based electronic signatures use physical devices such as USB tokens, smart cards, or SIM cards to store cryptographic keys, which must be securely managed following key management standards such as NIST SP 800-57~\cite{nist80057}. Authentication is based on possession of the device and the entry of a PIN / password.

\textbf{Advantages:}
\begin{itemize}
    \item High security: Requires both physical possession and user authentication.
    \item Strong legal validity: Recognized as QES in most legal frameworks.
\end{itemize}

\textbf{Disadvantages:}
\begin{itemize}
    \item Requires setup: The installation of software and drivers is necessary.
    \item Device dependency: If the token is lost, the sign-in cannot be performed.
\end{itemize}

\subsubsection{Acquisition Process}
\begin{enumerate}
    \item \textbf{Application}: Users apply through an authorized electronic certificate provider.
    \item \textbf{Identity Verification}: Users verify their identity physically at a certification authority’s office or via a notary.
    \item \textbf{Device Issuance}: A USB token or smart card is issued to the user.
    \item \textbf{PIN and Certificate Activation}: Users connect the device to their computer and set up their PIN.
\end{enumerate}

\subsubsection{Usage Process}
\begin{enumerate}
    \item \textbf{Signing}: Users connect the USB token or smart card, open the signing software, and enter their PIN to sign documents.
    \item \textbf{Verification}: Signed documents can be verified via Adobe Trust Center or eIDAS-compliant systems.
    \item \textbf{Device Management}: Users must reset their PIN if forgotten and renew certificates upon expiration.
\end{enumerate}

\section{Usability}

Usability is regarded as one of the most crucial components of quality for every type of product \cite{7023887}. Usability is regarded as one of the most crucial components of quality for every type of product~\cite{7023887, iso9241, nielsen1994}.
The concept of usability is applicable to a variety of product types. Testing the usability of hardware and software products is a growing trend as the subject of usability engineering gains popularity every day\cite{5608809}. Usability, in the context of e-signature systems, is essential not only for ensuring user satisfaction but also for promoting wider adoption. The complexity of these systems can lead to user frustration, errors, and potential security vulnerabilities. As electronic signatures are increasingly used for both personal and professional purposes, understanding the factors that influence their usability becomes crucial for improving their design and user experience. Previous research has highlighted that while many users perceive e-signatures as secure, there is often a disconnect between user perceptions and actual usability \cite{cagal}. This paper aims to explore the usability of two commonly used types of e-signatures—token-based and remote e-signatures—by evaluating their acquisition, installation, and usage from a user experience perspective.

\section{Related Work}

The usability and adoption of electronic signatures, particularly Qualified Electronic Signatures (QES), have been explored in several studies. However, existing research predominantly focuses on technical, legal, or security aspects rather than end-user usability and comparative experiences between token-based and remote signing methods.

Cagal and Bicakci \cite{cagal} analyzed the usability of Qualified Electronic Signatures (QES) by categorizing system designs and usage scenarios across Turkey and the European Union. Their study used cognitive walkthroughs to identify usability barriers in different QES implementations. While their work highlights system design flaws, it does not directly compare different QES methods (e.g., token-based vs. remote) in an experimental setting, which our study addresses.

Wang \cite{WANG200732} examined the legal frameworks of electronic signatures across countries such as the United States, United Kingdom, Germany, and China. The study emphasized that legal inconsistencies hinder cross-border digital signature adoption. Although Wang's study is crucial for understanding regulatory challenges, it does not evaluate usability from an end-user perspective. Our research complements this by focusing on user experience rather than legal interoperability.

Truong and Minh-Tuan \cite{truong} studied the use of digital signatures supported by Hardware Security Modules (HSM) in Vietnam’s e-invoicing system. Their focus was mainly on enhancing security in a specific industry context. However, their study did not investigate usability challenges faced by end-users across different signature methods, which is the primary aim of our study.

Radka et al. \cite{radka} evaluated the implementation of the eIDAS regulation in the Czech Republic, particularly the legal and procedural aspects of QES adoption. Although their work sheds light on regulatory inconsistencies within the EU, it does not analyze the impact of these regulations on user experience or preferences between signature methods.

Lax et al. \cite{Lax01072015} discussed vulnerabilities in digital document signing systems and proposed solutions to enhance system robustness. While their work is significant for system designers, it does not address the user's practical experiences during acquisition, installation, and usage phases of e-signature systems, which our study explores.

Paz and Pow-Sang \cite{7023887} conducted a systematic review on usability evaluation methods for software products. Their findings stress the importance of integrating usability engineering into system design. However, they do not specifically apply these principles to the context of electronic signatures, leaving a gap that our study aims to fill by applying usability evaluation specifically to token-based and remote e-signature systems.

Last et al. \cite{last2024} designed and evaluated a prototype for signing digital documents using digital identity wallets. Their study focuses on improving the security and intuitiveness of signing processes through verified personal attributes, but does not directly compare different QES systems from a usability perspective as our study does.

ENISA (European Union Agency for Cybersecurity) \cite{enisa2017} provided security guidelines for the appropriate use of qualified electronic registered delivery services. Although this guideline focuses on improving trust and interoperability in electronic communications, it does not address the comparative usability challenges between token-based and remote e-signatures that our study investigates.

\textbf{Research Gap:}  
While previous studies have addressed regulatory frameworks, security aspects, specific system designs, and general usability principles, there is a lack of empirical research comparing the usability and security perceptions of token-based and remote Qualified Electronic Signature systems through user-centered experiments. Our study addresses this gap by conducting a direct comparative usability evaluation with real users, analyzing acquisition, installation, usage experiences, and security perceptions systematically.

\section{Methodology}

To investigate the usability of token-based and remote electronic signatures, a user-centered study was conducted with 20 participants. The study aimed to evaluate the acquisition, installation, and usage processes of these two e-signature methods, with a particular focus on user experience and security perceptions. Participants were selected based on their diverse professional backgrounds, which allowed for a comprehensive analysis of usability across different user groups.

The study was organized into three main phases: the acquisition phase, the installation phase, and the usage phase. Each phase was assessed through surveys designed to capture participants' experiences, preferences, and security perceptions.

\subsection{Ethical Considerations}

This study involving human participants was reviewed and approved by the Social and Human Sciences Scientific Research and Publication Ethics Board/Istanbul Technical University. The approval reference number is 571 dated 04 November 2024. All participants provided informed consent prior to their participation. Participants were assured that their responses would be anonymized, participation was voluntary, and they could withdraw from the study at any time without consequences.

\subsection{Preliminary Survey: Participant Background and Expectations}
Before engaging in the usability study, participants completed a preliminary survey to collect demographic information and assess their prior experience with electronic signatures. The survey included the following key questions:
\begin{itemize}
    \item Age range (19-24, 25-29, 30-34, 35-39, 40-44, 45+)
    \item Gender (Male/Female)
    \item Education level (High School, Associate’s, Bachelor’s, Master’s, Doctorate)
    \item Prior experience with electronic signatures (Yes/No)
    \item Type of electronic signature previously used (Remote, Token-based, Mobile, ID card-based)
    \item Satisfaction level with prior e-signature usage (1: Not satisfied at all - 5: Very satisfied)
    \item Self-reported proficiency in using computers and mobile devices (1: Not competent at all - 5: Very competent)
    \item Previous experience with digital identity verification (e.g., e-Government, banking authentication, etc.)
    \item Expected ease of e-signature acquisition (1: Should be very difficult - 5: Should be very easy)
    \item Open-ended question: If you have used an e-signature before, please describe your experience.
\end{itemize}
These responses provided valuable context regarding participants' expectations and familiarity with digital authentication processes.

\subsection{Phase 1: Acquisition of E-Signatures}
During the acquisition phase, participants were not required to individually register for a new token-based or remote e-signature system. Instead, electronic signatures that had been previously obtained by the researchers were used for the study. 

This decision was made to protect participants' personal data and to avoid imposing financial costs on them, as acquiring an electronic signature would have required participants to share sensitive personal information with third-party providers and to cover the associated acquisition fees.

The acquisition processes were explained to the participants in detail by the researchers, simulating the key steps typically involved in real-world acquisition procedures. Specifically, the identity verification phase was reenacted by the researchers to demonstrate the authentication steps required for each e-signature method. Participants were informed about standard procedures, including application, identity verification, and credential activation, but were not required to perform these actions themselves.

Following the simulated acquisition process, participants completed an acquisition survey in which they rated statements on a 5-point Likert scale (1: Strongly Disagree - 5: Strongly Agree):
\begin{itemize}
    \item Preference for an online acquisition process
    \item Perceived ease of acquiring a token-based e-signature
    \item Perceived ease of acquiring a remote e-signature
    \item Perceived security of the identity verification process
    \item Technical issues that would be expected during acquisition
\end{itemize}
Additionally, participants selected their preferred acquisition method based on the simulation and provided qualitative feedback on their impressions of the process.

\subsection{Phase 2: Installation and Setup}
In the installation phase, participants were required to install the necessary software for the token-based e-signature system, while remote e-signature users set up their profiles for signing documents online. The installation process was carefully observed to identify any issues related to system compatibility, user errors, and overall ease of installation.

Participants completed an installation survey, where they rated the following statements:
\begin{itemize}
    \item Ease of installing the token-based e-signature software
    \item Clarity of installation instructions
    \item Preference for an e-signature system that does not require installation
\end{itemize}
Participants were also asked to choose their preferred installation method and explain their reasoning.

\subsection{Phase 3: Usage and Security Perceptions}
The usage phase involved participants signing documents using both the token-based and remote e-signatures. The signing process was evaluated based on factors such as ease of use, time taken to complete the task, the clarity of the interface, and any technical issues encountered.

Participants rated their experiences with both methods using the following criteria:
\begin{itemize}
    \item Ease of signing documents using a token-based e-signature
    \item Speed of the token-based signing process
    \item Absence of technical issues with the token-based system
    \item Perceived security of token-based e-signatures
    \item Ease of signing documents using a remote e-signature
    \item Speed of the remote signing process
    \item Absence of technical issues with the remote system
    \item Perceived security of remote e-signatures
\end{itemize}
Participants also indicated which method they found more secure and more usable, providing justifications for their choices.

\subsection{Task Design and Survey Mapping}

In order to comprehensively assess the usability and security perceptions of token-based and remote electronic signatures, participants were assigned a series of tasks during the study. Each task was carefully designed to simulate real-world scenarios and was followed by targeted surveys to capture user feedback and experiences.

\begin{itemize}
  \item \textbf{Task 1: Observation of E-Signature Acquisition Processes}

Participants were provided with a detailed walkthrough of the acquisition processes for both remote and token-based electronic signatures. The researchers explained each step of the acquisition procedures, including application submission, identity verification, and certificate activation. As part of the simulation, participants were shown examples of the actual online forms required for the acquisition of remote and token-based electronic signatures. The process was demonstrated through screenshots of real-world form submission interfaces.

\begin{figure}[h]
    \centering
    \includegraphics[width=0.9\textwidth]{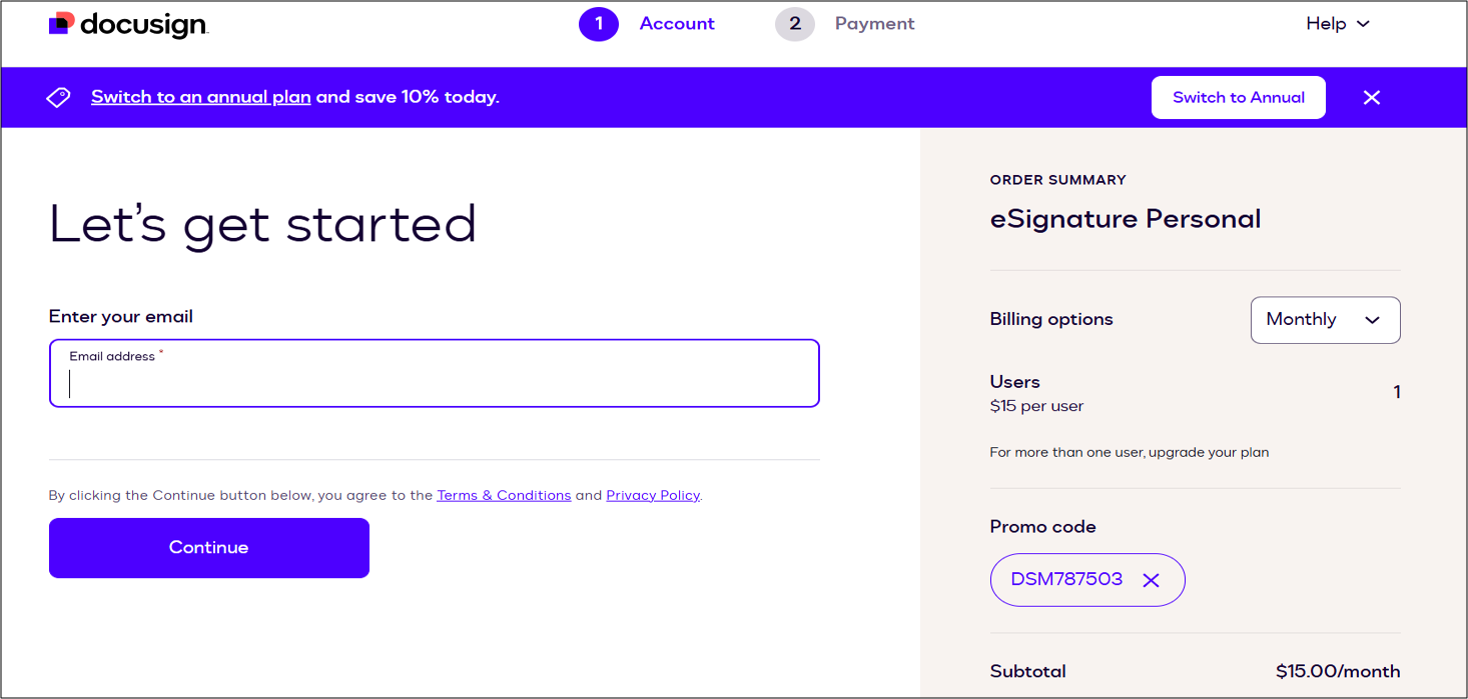}
    \caption{Example of an Online Application Form for Remote E-Signature Acquisition}
    \label{fig:remote_form}
\end{figure}

\begin{figure}[h]
    \centering
    \includegraphics[width=0.9\textwidth]{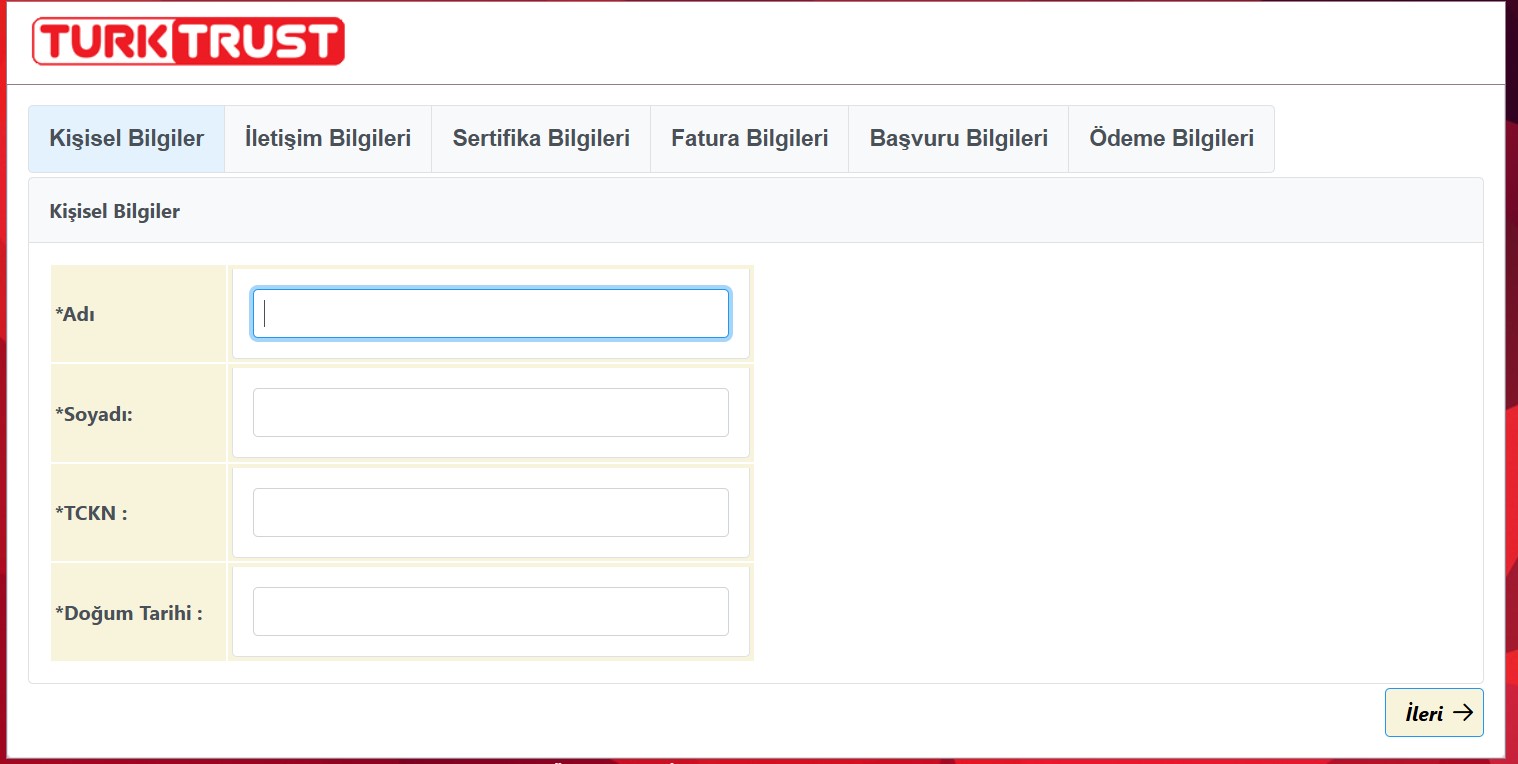}
    \caption{Example of an Online Application Form for Token-Based E-Signature Acquisition}
    \label{fig:token_form}
\end{figure}

The screenshots helped participants visualize the required data fields and steps involved in obtaining an e-signature, making the simulation more realistic. After the demonstration, participants were asked to complete the \textit{Acquisition Survey} to evaluate the perceived ease, security, and potential difficulties associated with the acquisition phase.

\textit{Associated Survey: Acquisition Survey}

  \item \textbf{Task 2: Token-Based E-Signature Software Installation}

Participants were instructed to install the token-based e-signature software using the installation guide provided by the e-signature service provider. The guide detailed two major steps: the installation of the Palma software and the activation of the token.

\begin{figure}[h]
    \centering
    \includegraphics[width=0.9\textwidth]{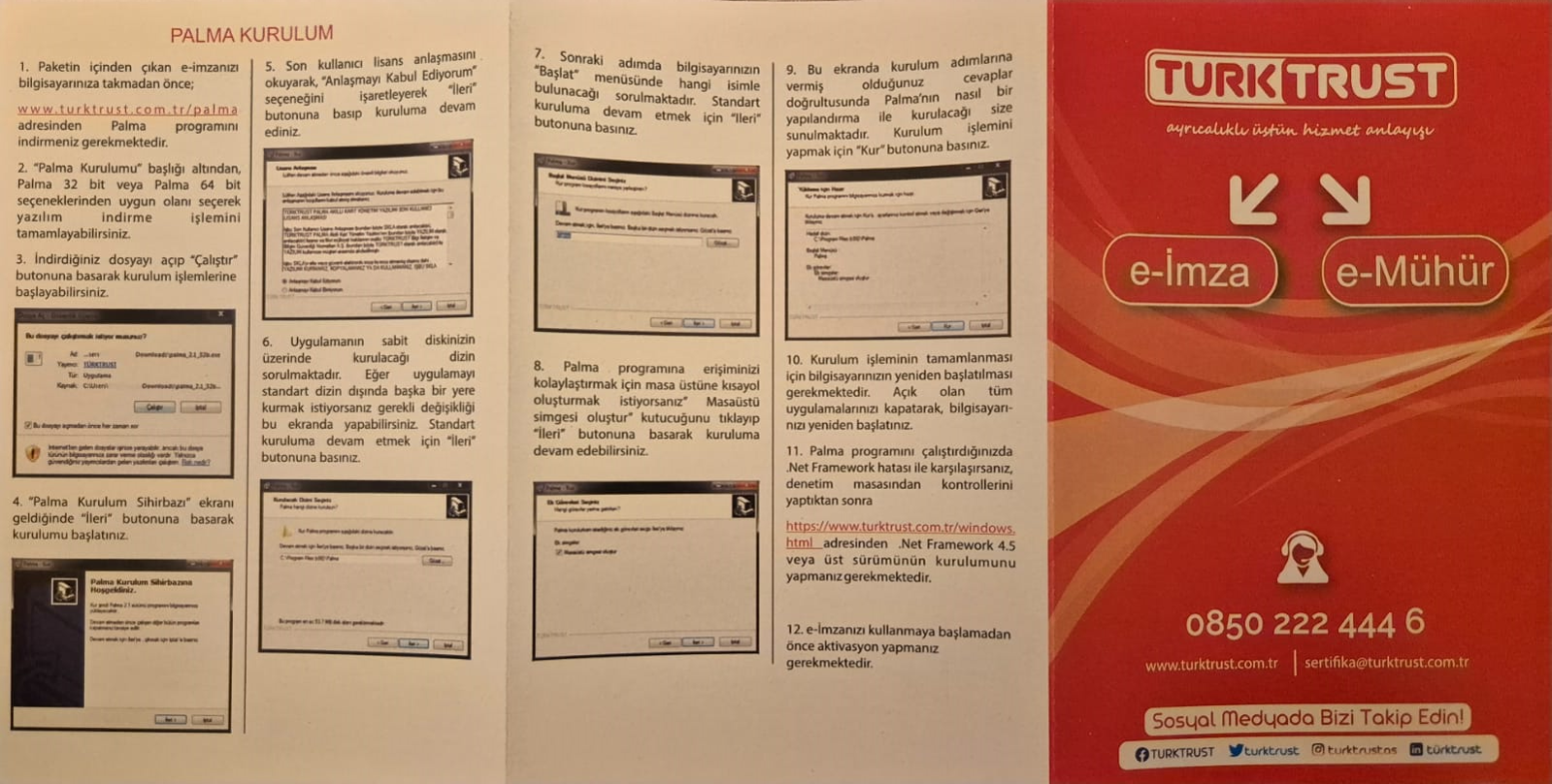}
    \caption{Installation Steps for Palma Software Required for Token-Based E-Signature}
    \label{fig:token_installation}
\end{figure}

Following the installation, participants were guided through the activation process, which included setting up the service agreement and configuring the necessary authentication settings.

\begin{figure}[h]
    \centering
    \includegraphics[width=0.9\textwidth]{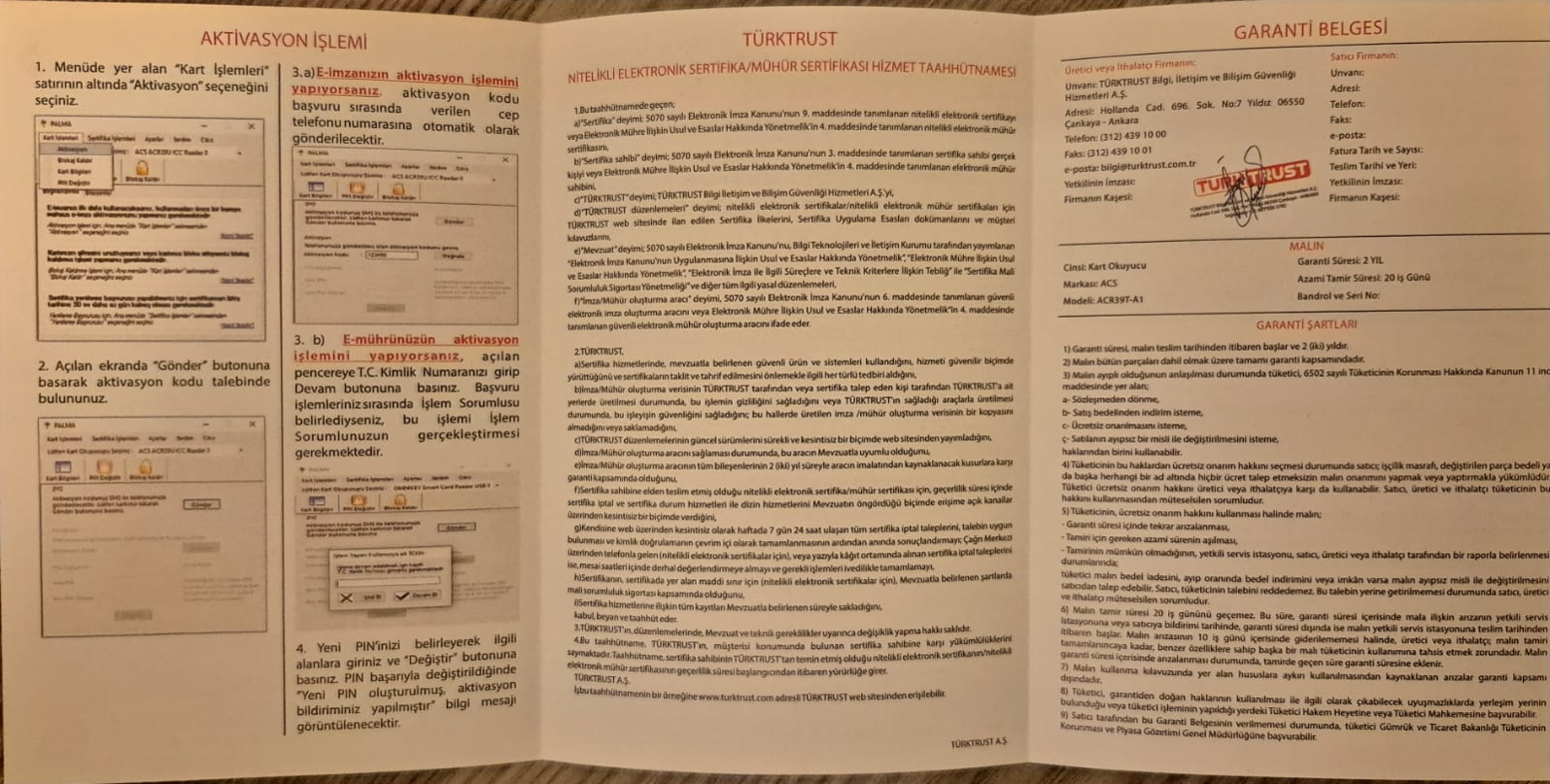}
    \caption{Activation Process and Service Agreement for Token-Based E-Signature}
    \label{fig:token_activation}
\end{figure}

Participants were observed during the installation and activation tasks. Any technical difficulties or support needs were recorded by the researchers. Participants subsequently completed the \textit{Installation Survey} to evaluate their experiences.

\textit{Associated Survey: Installation Survey}

    \item \textbf{Task 3: Signing a Document with Token-Based E-Signature}

Participants were tasked with signing a PDF document using a USB token device. They connected the token to their computers, launched the signature application (Palma), entered their PIN codes, and completed the signing process.

\begin{figure}[h]
    \centering
    \includegraphics[width=0.7\textwidth]{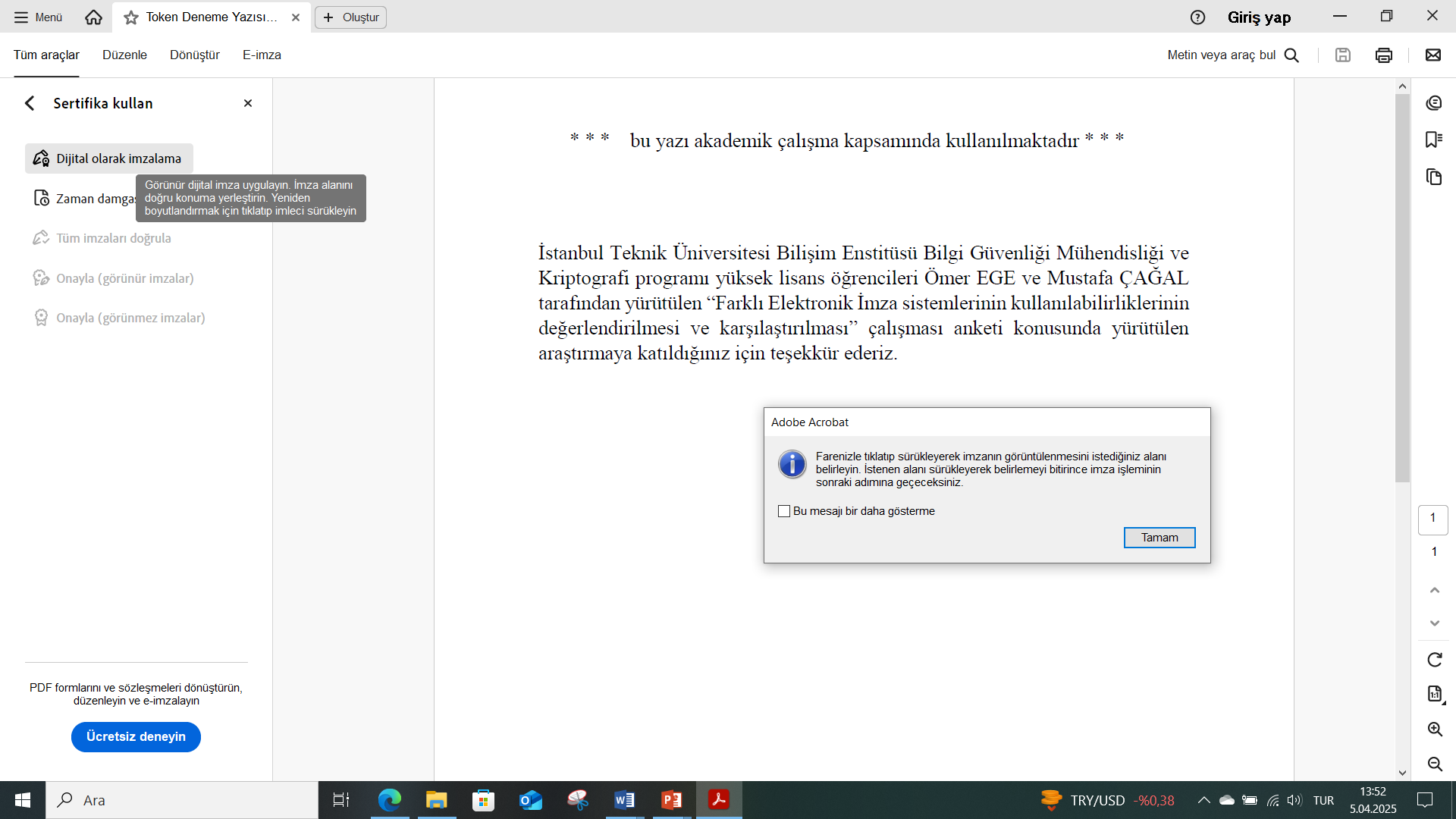}
    \caption{Interface for Signing a Document Using a Token-Based E-Signature}
    \label{fig:token_signing}
\end{figure}

During this task, participants were observed for any difficulties related to accessing the token device, entering the PIN, or interacting with the signing software. Participants subsequently evaluated their experience by completing the \textit{Token-Based Usage Survey}.

\textit{Associated Survey: Token-Based Usage Survey}

   \item \textbf{Task 4: Verification of a Document Signed with Token-Based E-Signature}

Participants were asked to verify the authenticity of a document they had signed using the token-based e-signature. They used Adobe Reader's signature verification functionality to check the validity of the electronic signature.

\begin{figure}[h]
    \centering
    \includegraphics[width=0.7\textwidth]{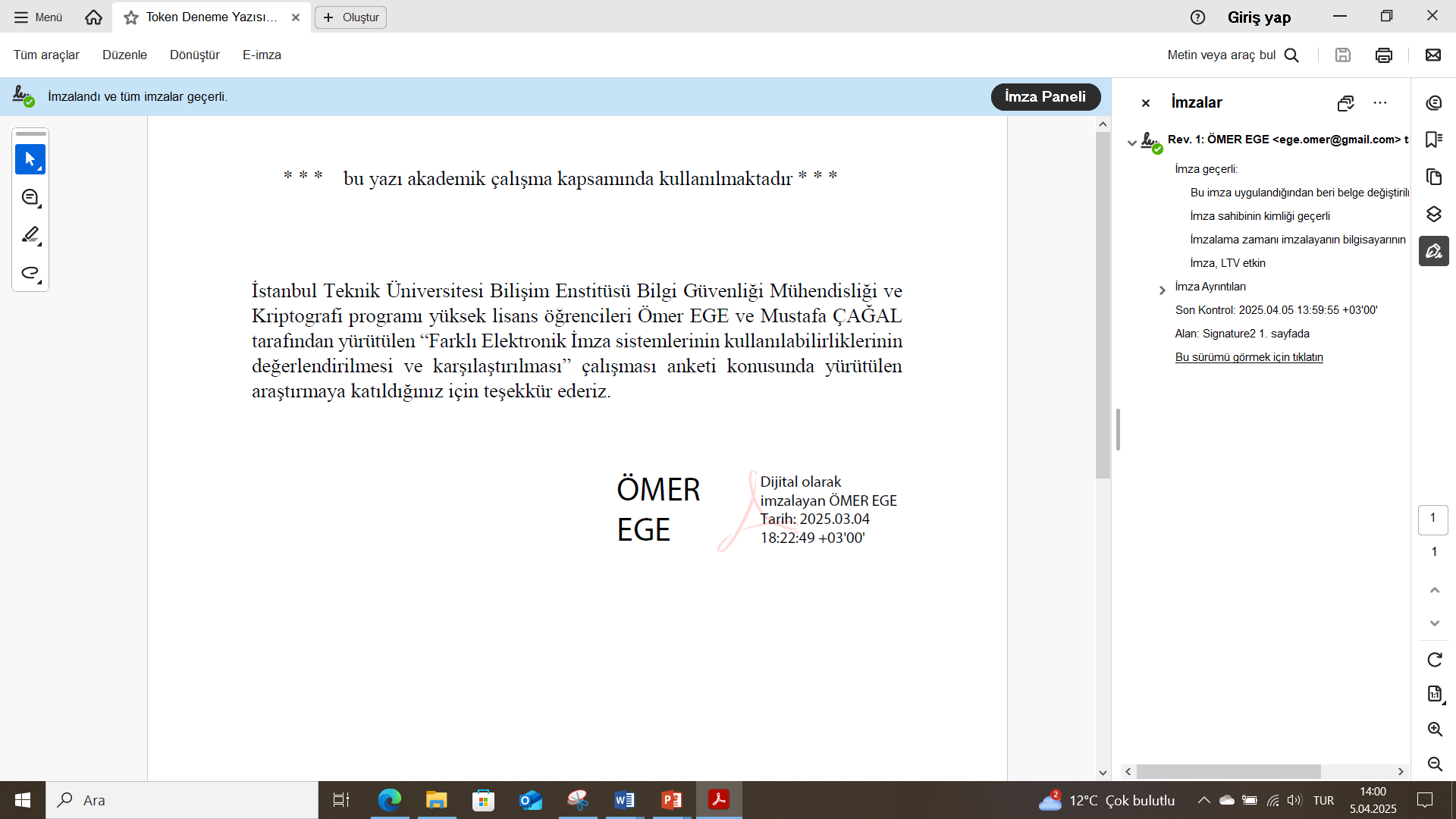}
    \caption{Verification of a Token-Based Signed Document in Adobe Reader}
    \label{fig:token_verification}
\end{figure}

Participants were observed while performing the verification, and any difficulties in understanding verification indicators or system messages were noted.

\textit{Associated Survey: Token-Based Usage Survey (Verification Section)}

   \item \textbf{Task 5: Signing a Document with Remote E-Signature}

Participants signed a PDF document using a web-based remote signature platform (DocuSign). They logged into the platform, uploaded a document, and completed the signing process by verifying their identity via a One-Time Password (OTP) sent to their registered mobile number.

\begin{figure}[h]
    \centering
    \begin{subfigure}[b]{0.3\textwidth}
        \includegraphics[width=\textwidth]{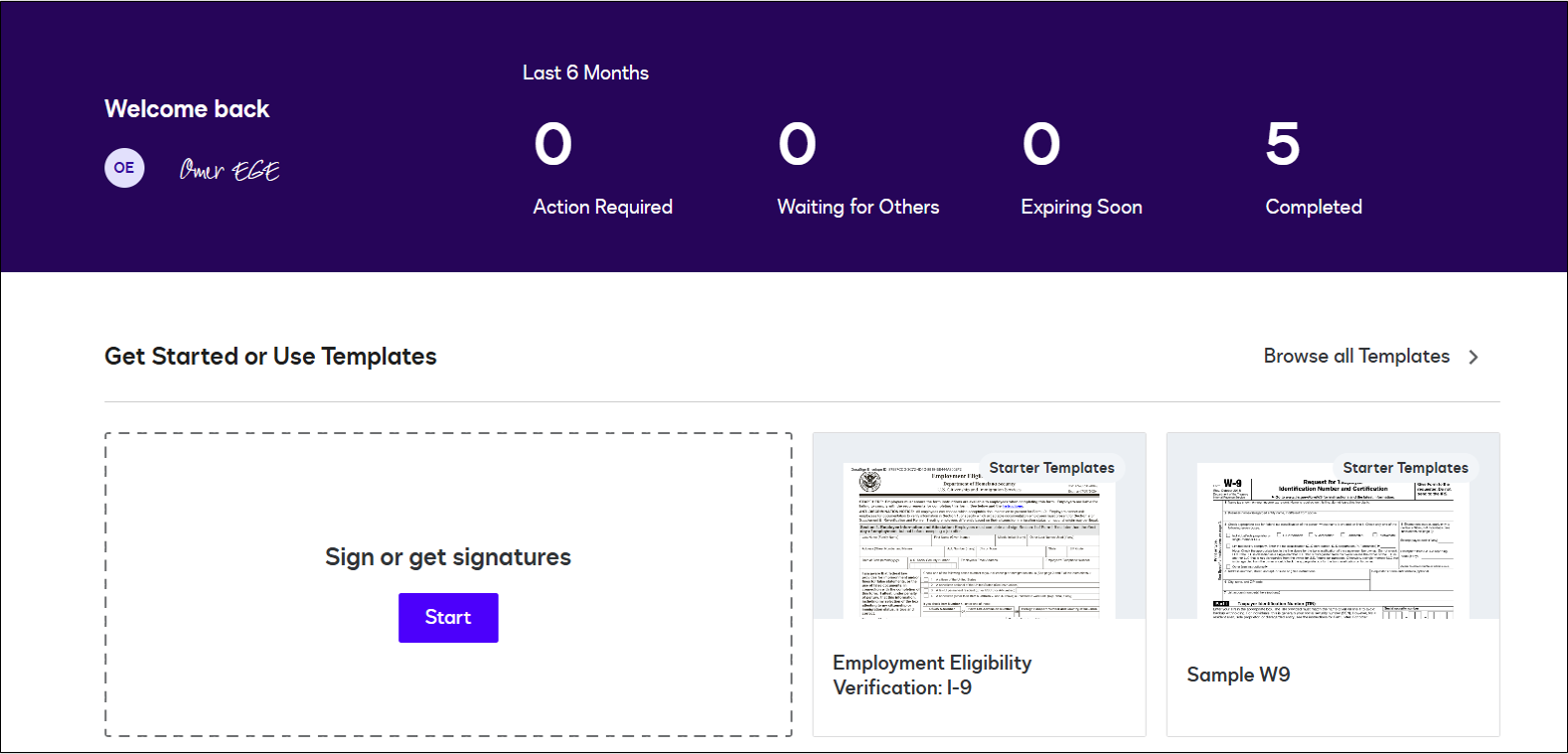}
        \caption{Step 1: Login to the Platform}
        \label{fig:remote_step1}
    \end{subfigure}
    \hfill
    \begin{subfigure}[b]{0.3\textwidth}
        \includegraphics[width=\textwidth]{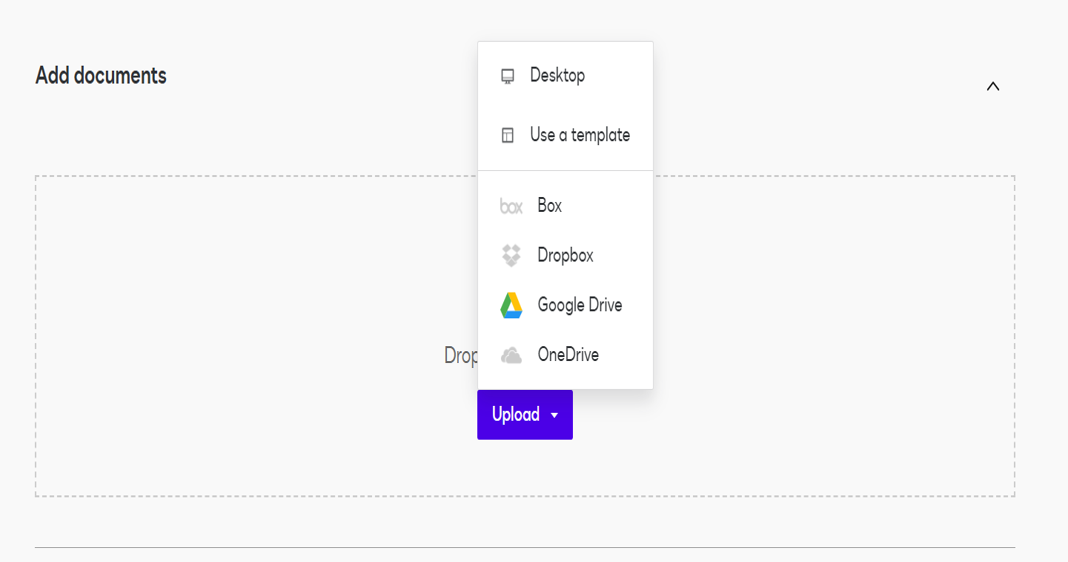}
        \caption{Step 2: Upload the Document}
        \label{fig:remote_step2}
    \end{subfigure}
    \hfill
    \begin{subfigure}[b]{0.3\textwidth}
        \includegraphics[width=\textwidth]{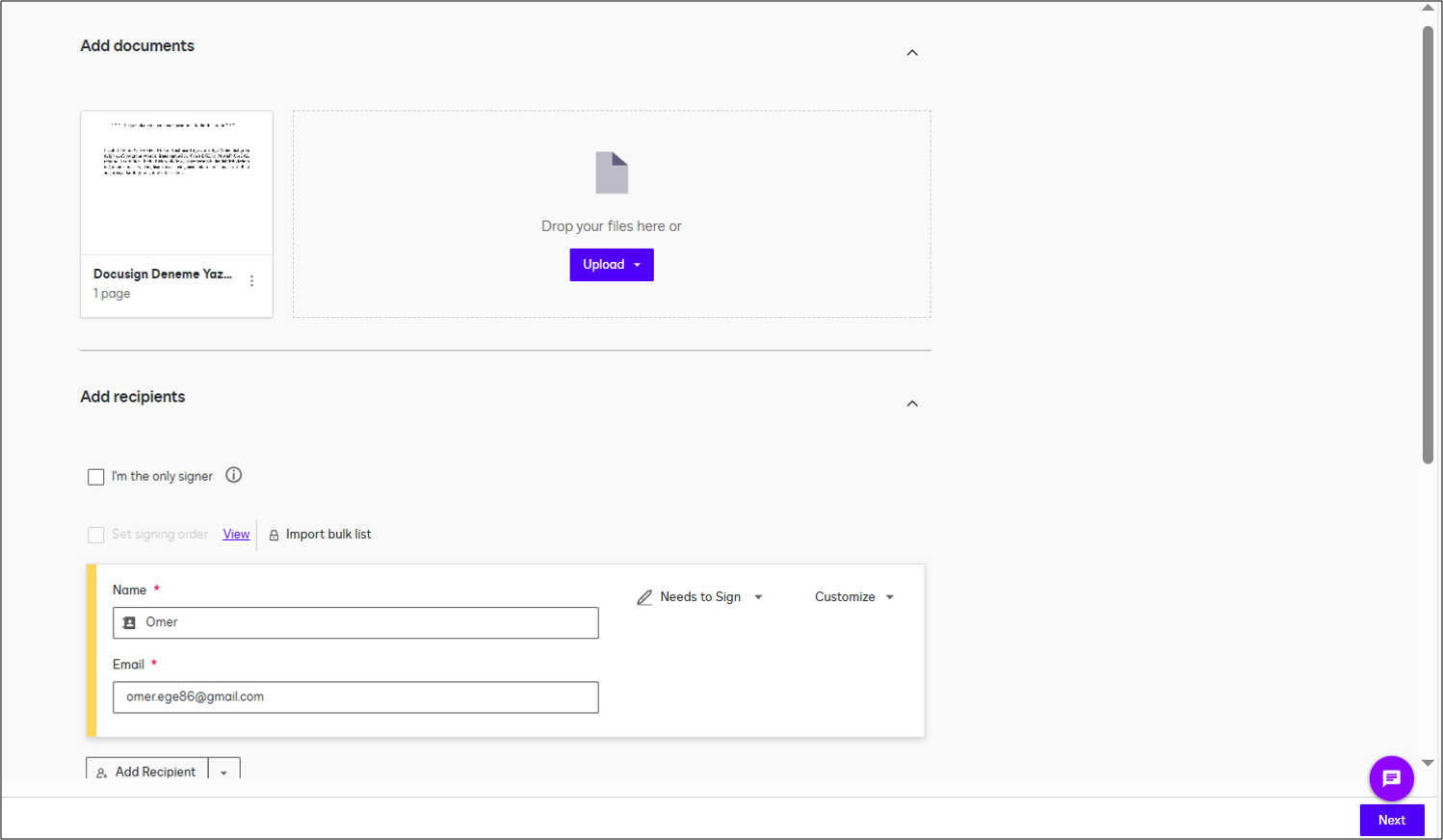}
        \caption{Step 3: Add Recipients}
        \label{fig:remote_step3}
    \end{subfigure}

    \vspace{0.5cm}

    \begin{subfigure}[b]{0.3\textwidth}
        \includegraphics[width=\textwidth]{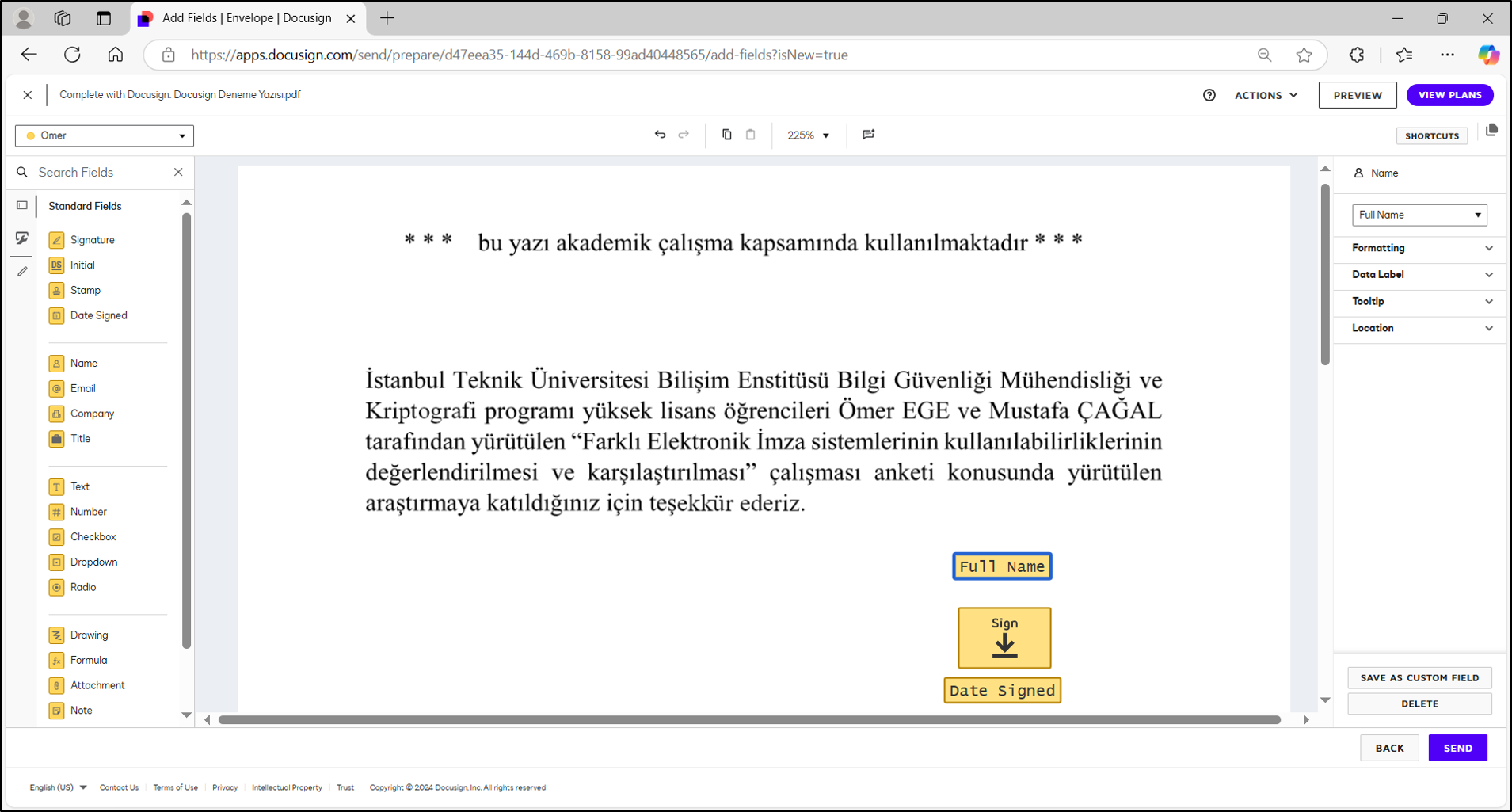}
        \caption{Step 4: Place Signature Field}
        \label{fig:remote_step4}
    \end{subfigure}
    \hfill
    \begin{subfigure}[b]{0.3\textwidth}
        \includegraphics[width=\textwidth]{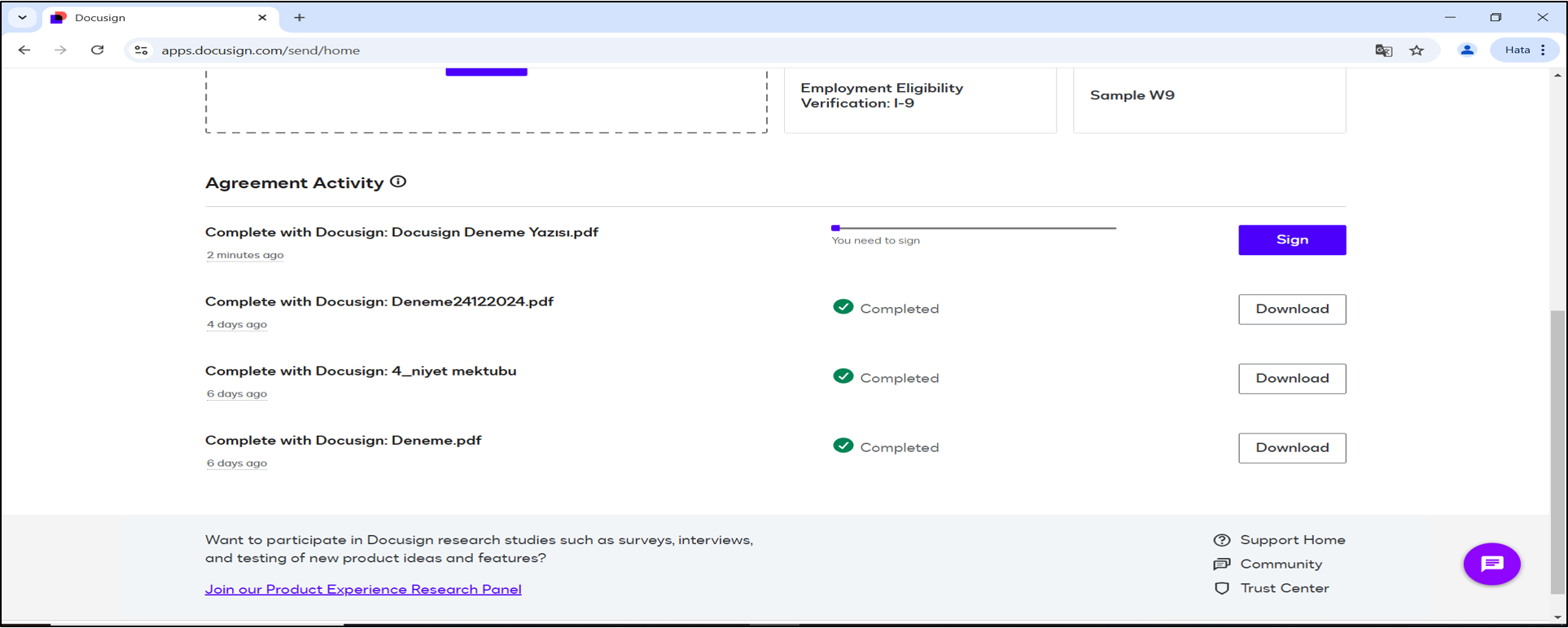}
        \caption{Step 5: Check the status}
        \label{fig:remote_step5}
    \end{subfigure}
    \hfill
    \begin{subfigure}[b]{0.3\textwidth}
        \includegraphics[width=\textwidth]{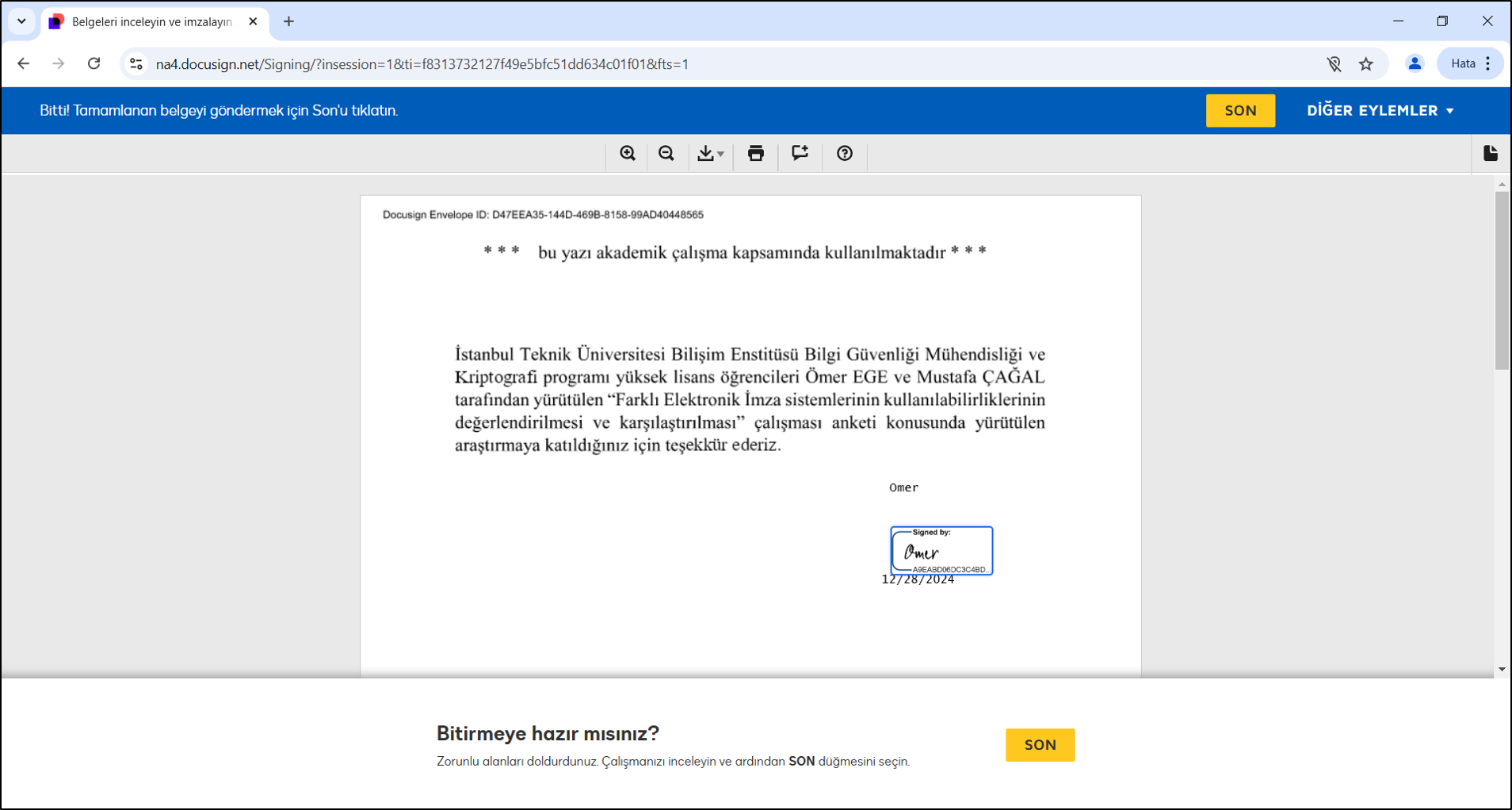}
        \caption{Step 6: Complete the Signature}
        \label{fig:remote_step6}
    \end{subfigure}

    \caption{Remote E-Signature Signing Steps (1–6)}
    \label{fig:remote_signing_steps1}
\end{figure}

Participants were observed for usability issues such as difficulties in navigating the platform, uploading files, or completing the OTP verification. Participants subsequently completed the \textit{Remote Usage Survey}.

\textit{Associated Survey: Remote Usage Survey}

   \item \textbf{Task 6: Verification of a Document Signed with Remote E-Signature}

Participants verified the authenticity of a document signed via the remote signature platform using the platform’s built-in verification tool. This process included checking the digital certificate details and ensuring the document integrity.

\begin{figure}[h]
    \centering
    \includegraphics[width=0.7\textwidth]{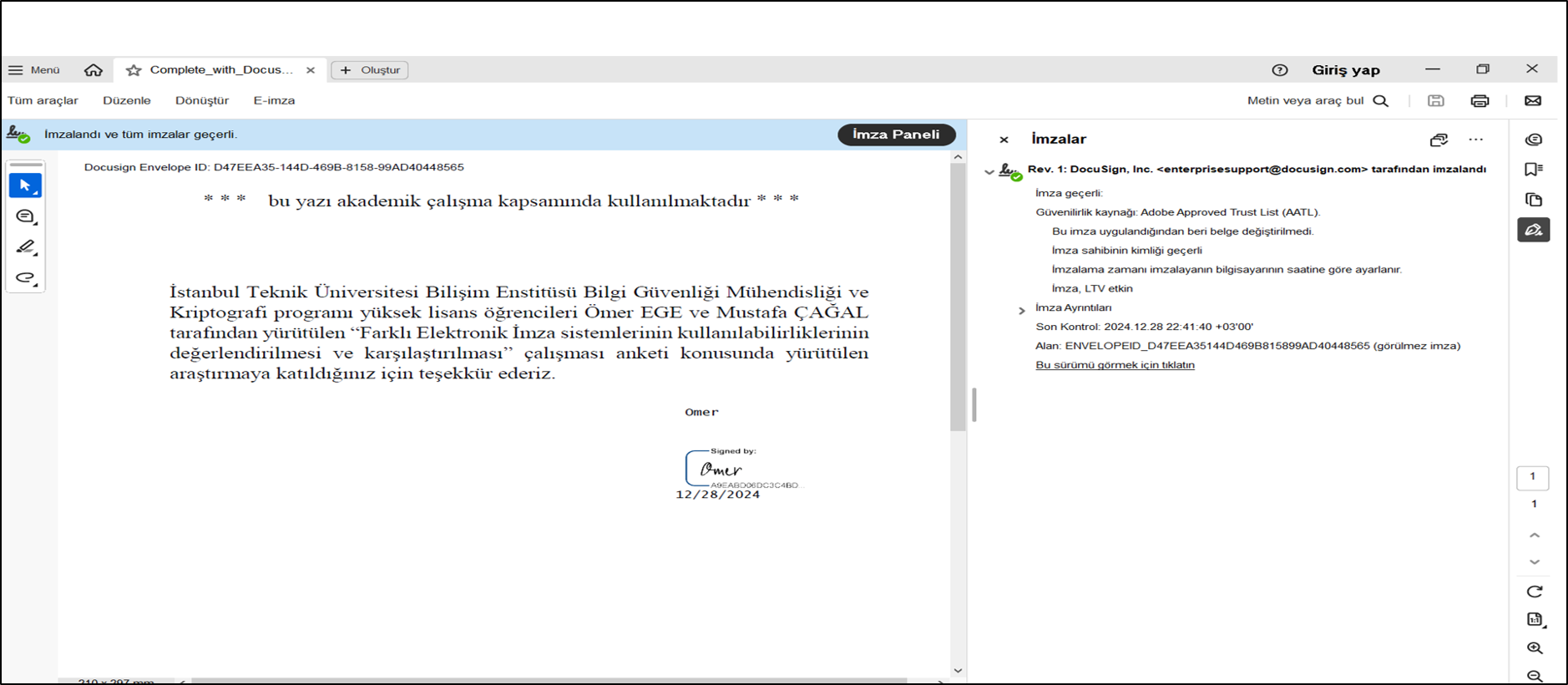}
    \caption{Verification of a Remote Signed Document on the Web Platform}
    \label{fig:remote_verification}
\end{figure}

Participants were observed while reviewing the verification information, and any misunderstandings or difficulties were recorded.

\textit{Associated Survey: Remote Usage Survey (Verification Section)}

\end{itemize}

At the conclusion of all tasks, participants completed a final comparative survey where they evaluated both signature methods based on usability, perceived security, installation complexity, and overall preference. Qualitative feedback was also collected to capture deeper insights into user experiences and expectations.

\subsection{Data Collection and Analysis}
Data were collected through surveys completed by the participants at each phase of the study. These surveys included Likert-scale questions to assess the participants’ satisfaction with the acquisition, installation, and usage processes. Participants were also asked to compare the two e-signature methods in terms of usability, security, and overall preference.

The results were analyzed using descriptive statistics and comparative analysis. Usability scores, error rates, and participant feedback were compared across both e-signature methods to determine which system provided a better overall user experience. Additionally, qualitative responses were examined to identify recurring themes and key concerns. This analysis aimed to contribute to the development of more user-friendly and secure e-signature systems that meet the needs of a wide range of users.

\subsection{Hypotheses}
Based on prior research and preliminary observations, the following hypotheses were formulated:

\begin{itemize}
    \item \textbf{H1:} Remote electronic signatures will be perceived as more usable than token-based electronic signatures.
    \item \textbf{H2:} Token-based electronic signatures will be perceived as more secure than remote electronic signatures.
    \item \textbf{H3:} Participants will prefer remote electronic signatures over token-based signatures due to ease of acquisition and usage.
\end{itemize}

\section{Results}

This section presents the findings obtained from the user study conducted to evaluate the usability of token-based and remote electronic signatures. The results are categorized into three key areas: acquisition, installation, and usage, followed by an analysis of security perceptions and user preferences.

\subsection{Participant Demographics}
The study included 20 participants aged between 19 and 45+. Among them, 75\% were male and 25\% were female. Regarding educational background, 5\% had an associate's degree, 65\% held a bachelor's degree, 15\% had a master's degree, and 15\% had a doctorate.

\begin{table}[h]
\centering
\caption{Participant Demographics}
\resizebox{\textwidth}{!}{%
\begin{tabular}{|l|c|}
\hline
\textbf{Demographic Category} & \textbf{Percentage (\%)} \\
\hline
Age 19–24 & 10.0 \\
Age 25–29 & 10.0 \\
Age 35–39 & 35.0 \\
Age 40–44 & 20.0 \\
Age 45+ & 25.0 \\
\hline
Male & 75.0 \\
Female & 25.0 \\
\hline
Associate's Degree & 5.0 \\
Bachelor's Degree & 65.0 \\
Master's Degree & 15.0 \\
Doctorate & 15.0 \\
\hline
\end{tabular}%
}
\label{tab:demographics}
\end{table}

\subsection{Acquisition Phase}
Participants evaluated the ease of acquiring both token-based and remote e-signatures. The findings indicate that \textbf{75.0\% of participants found the remote e-signature acquisition process easier} compared to token-based signatures. This preference is attributed to the fully online nature of the remote e-signature system, which eliminates the need for physical visits to a service provider or dealing with hardware tokens.

Furthermore, \textbf{75.0\% of participants preferred an entirely online acquisition process}, highlighting the growing importance of digital accessibility and convenience in modern workflows. 

The security perception during acquisition remained high for both methods, with an average rating of \textbf{4.7 out of 5} for the identity verification process.

\subsection{Installation Phase}
Installation complexity remained a significant factor influencing user experience. The findings reveal that \textbf{85.0\% of participants preferred remote e-signatures} due to the absence of installation requirements. In contrast, token-based e-signatures necessitated software installation and configuration, which led to technical difficulties for many users.

The average difficulty rating for the token-based installation process was \textbf{3.2 out of 5}, with participants frequently reporting issues related to system compatibility, driver installation, and dependencies on external software such as Java or middleware applications. These challenges often resulted in increased setup time and reduced user satisfaction compared to the remote e-signature method.

\subsection{Usage Phase}
In terms of practical usage, \textbf{90.0\% of participants found the remote e-signature process more accessible and user-friendly}. The simplicity of signing documents without requiring additional hardware was a key factor in this preference.

However, security perceptions varied between the two methods. \textbf{65.0\% of participants considered token-based e-signatures more secure}, citing the need for a physical device as an additional layer of protection. Conversely, some participants expressed concerns about remote e-signatures, particularly regarding password-based authentication and the potential risk of unauthorized access.

\begin{table}[h]
    \centering
    \caption{Survey Results for Usability and Security Perceptions}
    \resizebox{\textwidth}{!}{
    \begin{tabular}{|l|c|c|}
        \hline
        \textbf{Survey Question} & \textbf{Remote E-Signature (\%)} & \textbf{Token-Based E-Signature (\%)} \\
        \hline
        Found the acquisition process easy & 75.0 & 25.0 \\
        Preferred an entirely online acquisition process & 75.0 & 25.0 \\
        Rated identity verification security (Avg. 1-5) & 4.7 & 4.55 \\
        Preferred installation process & 85.0 & 15.0 \\
        Found installation process difficult (Avg. 1-5) & 2.0 & 3.2 \\
        Found the usage process easy & 90.0 & 10.0 \\
        Considered the method secure & 20.0 & 80.0 \\
        Preferred method for signing documents & 70.0 & 30.0 \\
        \hline
    \end{tabular}
    }
    \label{tab:survey_results}
\end{table}
\subsection{Task Performance Results}

The task completion rates and task durations for each phase of the study were recorded to evaluate the operational feasibility and participant engagement. Table~\ref{tab:task_performance} summarizes the success rates and average time spent per task.

\begin{table}[h]
\centering
\caption{Task Completion Rates and Average Time Measurements}
\label{tab:task_performance}
\resizebox{\textwidth}{!}{%
\begin{tabular}{|l|c|c|c|}
\hline
\textbf{Task} & \textbf{Completed Successfully (\%)} & \textbf{Failed (\%)} & \textbf{Average Time (minutes)} \\
\hline
Task 1: Observation of E-Signature Acquisition Processes & 100.0 & 0.0 & N/A \\
Task 2: Token-Based E-Signature Software Installation & 55.0 & 45.0 & 14.0 \\
Task 3: Signing a Document with Token-Based E-Signature & 100.0 & 0.0 & N/A \\
Task 4: Verification of a Document Signed with Token-Based E-Signature & 100.0 & 0.0 & N/A \\
Task 5: Signing a Document with Remote E-Signature & 100.0 & 0.0 & N/A \\
Task 6: Verification of a Document Signed with Remote E-Signature & 100.0 & 0.0 & N/A \\
\hline
\end{tabular}%
}
\end{table}

\noindent
\textbf{Analysis:}

\begin{itemize}
    \item \textbf{Task 1:} All participants successfully completed the observation of acquisition processes.
    
\item \textbf{Task 2:} In the installation task, \textbf{55.0\% of participants} successfully installed the token-based e-signature software without major issues, whereas \textbf{45.0\%} encountered various technical difficulties. 

The primary challenges reported during the installation phase included:
\begin{itemize}
    \item \textbf{System Compatibility Issues:} Several participants faced difficulties identifying whether their computer was 32-bit or 64-bit, causing confusion in selecting and installing the correct driver versions.
    
    \begin{quote}
    One participant noted: 
    \textit{``I had no idea whether my computer was 32 or 64-bit. I downloaded the wrong driver twice and kept getting errors until I figured it out by trial and error. It felt like something only IT professionals could set up properly.''}
    \end{quote}
    
    \item \textbf{Software and Middleware Requirements:} Some participants struggled with missing or outdated middleware components such as Java or specific token management applications (e.g., Palma), leading to installation errors and delays.
    \item \textbf{Incomplete Guidance:} Participants who lacked prior experience with token installations often needed to watch external instructional videos or rely on trial-and-error approaches, further prolonging the setup time.
    \item \textbf{Cumulative Troubleshooting:} Participants encountering multiple minor issues (e.g., browser settings, security permissions) experienced compounded frustration, significantly extending the time required for successful installation.
\end{itemize}

These challenges contributed to an average installation time of approximately \textbf{14 minutes}, which is relatively high for a standard software setup. This extended duration reflects the technical complexity of token-based systems, the variability in participants' device configurations, and the lack of streamlined, user-friendly installation processes.

\begin{quote}
One participant remarked: 
\textit{``I thought it would be a simple setup, but it turned out to be way more complicated than I expected.''}
\end{quote}

    \item \textbf{Task 3 \& 4:} All participants successfully signed and verified a document using the token-based e-signature. Minor technical issues such as delays in device recognition and occasional PIN entry errors were observed but did not prevent task completion.
    
    \item \textbf{Task 5 \& 6:} Similarly, all participants were able to successfully sign and verify a document using the remote e-signature platform. The remote signing process was generally perceived as more intuitive, requiring fewer technical interactions and significantly reducing user errors.
\end{itemize}

Overall, the results highlight that while token-based e-signature systems offer strong security advantages, the installation phase poses a significant barrier to usability, especially for users with lower technical proficiency or unfamiliarity with system-level configurations.

\subsection{User Preferences and Qualitative Insights}
When asked about their preferred e-signature method, \textbf{70.0\% of participants chose remote e-signatures}, whereas \textbf{30.0\% opted for token-based e-signatures}.

Participants who preferred remote e-signatures primarily cited ease of access, the absence of installation requirements, and the ability to sign documents across multiple devices without physical dependencies. Some notable comments included:
\begin{itemize}
    \item \textit{``The ability to use it anywhere without extra hardware is a major advantage.''} (Participant 8)
    \item \textit{``I prefer a solution that does not require software installation or additional configurations. Installing programs sometimes causes security risks.''} (Participant 13)
    \item \textit{``Web-based signing feels faster and more practical. I can sign documents from any device without worrying about compatibility.''} (Participant 3)
    \item \textit{``Carrying an extra device increases the risk of losing it. I feel more comfortable with a method that does not require carrying anything.''} (Participant 14)
    \item \textit{``Program installation can lead to malware risks. I trust browser-based platforms more.''} (Participant 17)
\end{itemize}

On the other hand, participants who preferred token-based e-signatures emphasized the perceived security benefits of physical devices. Their notable comments included:
\begin{itemize}
    \item \textit{``Having a physical device adds an extra layer of security. Even if my password is stolen, the attacker would still need the token.''} (Participant 2)
    \item \textit{``Token-based authentication provides more control over my electronic signature. I feel safer when a physical key is required.''} (Participant 6)
    \item \textit{``My signature key is not constantly exposed to the internet. That's why I prefer token-based solutions.''} (Participant 19)
    \item \textit{``Online platforms can be hacked. With a token, the risk is lower because it is in my possession.''} (Participant 9)
    \item \textit{``Even if remote signing is easier, I value security over convenience.''} (Participant 11)
\end{itemize}

Overall, the qualitative insights reflect a trade-off between \textbf{usability and convenience} in remote e-signatures versus \textbf{enhanced perceived security} in token-based solutions.

This suggests that reducing the installation complexity could significantly improve user adoption rates for token-based e-signature systems.

\subsection{Statistical Analysis of User Responses}

In order to test the study hypotheses, statistical analyses were conducted based on users' survey responses.

\subsubsection{Security Perception}

Users were asked which e-signature method they found more secure. A chi-square test for goodness of fit revealed a statistically significant difference, \(\chi^2(1, N=20) = 7.20, p < 0.01\). A large majority of users (16 out of 20) perceived token-based e-signatures as more secure than remote e-signatures, strongly supporting Hypothesis 2 (H2).

\subsubsection{Usability Perception}

Users were asked which e-signature method they found more usable. A chi-square test indicated a statistically significant difference, \(\chi^2(1, N=20) = 12.80, p < 0.001\). A clear majority of users (18 out of 20) found remote e-signatures to be more usable than token-based e-signatures, strongly supporting Hypothesis 1 (H1).

\subsubsection{Signing Method Preference}

Users were also asked which e-signature method they would prefer for signing documents. A one-sided binomial test was conducted to evaluate whether significantly more users preferred remote e-signatures over token-based ones. The test result was marginally above the conventional threshold for statistical significance, \(p = 0.058\), based on 14 out of 20 participants favoring the remote method. Although Hypothesis 3 (H3) was not fully supported, this finding indicates a notable trend toward significance, suggesting that convenience and ease of use may influence users’ preferences in real-world signing scenarios.

\begin{table}[h]
    \centering
    \caption{Summary of Statistical Tests}
    \begin{tabular}{|p{4cm}|p{3cm}|p{2cm}|p{3cm}|}
        \hline
        \textbf{Evaluation} & \textbf{Test} & \textbf{p-value} & \textbf{Result} \\
        \hline
        Security Perception & Chi-Square & \(<0.01\) & Significant \\
        Usability Perception & Chi-Square & \(<0.001\) & Significant \\
        Signing Preference & Binomial (One-sided) & 0.058 & Trend toward significance \\
        \hline
    \end{tabular}
    \label{tab:statistical_tests}
\end{table}

\section{Discussion}

The findings of this study highlight key usability differences between token-based and remote e-signatures. The strong preference for remote e-signatures aligns with previous research emphasizing the importance of accessibility and ease of use in digital authentication systems. 

\subsection{Hypotheses Evaluation}

The statistical analyses conducted support two out of the three initial hypotheses.

\begin{itemize}
    \item \textbf{H1} (Remote e-signatures are perceived as more usable) was strongly supported, as a significant chi-square result (\(p < 0.001\)) indicated that a large majority of users found remote e-signatures more usable than token-based ones.
    
    \item \textbf{H2} (Token-based e-signatures are perceived as more secure) was also supported, with a statistically significant difference (\(p < 0.01\)) showing that most users perceived token-based e-signatures as offering higher security.
    
    \item \textbf{H3} (Users prefer remote e-signatures for signing documents) was not supported, as a one-sided binomial test revealed that the observed preference for remote e-signatures (14 out of 20 participants) did not reach conventional statistical significance (\(p = 0.058\)). Although the result falls just short of the commonly used 0.05 threshold, it indicates a notable trend toward significance, suggesting that users may be inclined to favor remote e-signatures for signing tasks under certain conditions.
\end{itemize}

These results clearly demonstrate that users differentiate strongly between the two systems in terms of usability and perceived security. Remote e-signatures were found to be significantly more usable, likely due to their minimal setup requirements and ease of access. Token-based e-signatures, on the other hand, were perceived as more secure, reflecting users’ confidence in physical devices and hardware-backed protection. While the preference for remote e-signatures in signing tasks did not reach statistical significance, a notable trend toward significance was observed, suggesting that convenience may still play a critical role in user preferences. Taken together, these findings underscore the inherent trade-off between usability and security, which should inform the design and deployment of electronic signature systems.

\subsection*{Demographic Factors and Signing Method Preference}

Although the overall preference for remote e-signatures did not reach statistical significance, further exploratory analysis was conducted to examine how demographic variables—specifically education level, age group, and gender—might relate to signing method choices.

When grouped by education level, 5 out of 6 participants with postgraduate education (master’s or doctorate) preferred remote e-signatures. Among those with undergraduate education or lower, the distribution was more balanced, with 9 favoring remote and 5 choosing token-based signing. This pattern may reflect differences in digital confidence and trust models; more educated users may prioritize accessibility and efficiency, while those with less experience may rely on tangible security cues such as physical tokens.

Age-related patterns were also observed. Both the youngest (19–24) and oldest (45+) participants mostly preferred remote signatures, while the 35–39 group showed a more even split between the two methods. This may suggest that middle-aged users weigh security and convenience more equally, or that generational familiarity with digital systems plays a role in shaping preferences.

Gender-based trends indicated that male participants (n=15) were more likely to prefer remote methods (11 remote, 4 token), whereas female participants (n=5) were nearly evenly divided (3 remote, 2 token). Some female participants also expressed concerns about technical complexity or the potential loss of physical devices, reflecting a nuanced relationship between perceived usability, risk, and device trust.

While no statistical inference was drawn due to the limited sample size, these preliminary findings suggest that individual preferences in secure digital signing are shaped not only by system characteristics but also by users’ demographic backgrounds, security mindsets, and personal experiences with technology.

\begin{table}[h]
    \centering
    \caption{Signing Method Preference by Demographic Group}
    \begin{tabular}{|l|c|c|c|}
        \hline
        \textbf{Group} & \textbf{Remote Preference} & \textbf{Token Preference} & \textbf{Total} \\
        \hline
        \textbf{Education Level} & & & \\
        Undergraduate            & 9  & 5  & 14 \\
        Postgraduate             & 5  & 1  & 6 \\
        \hline
        \textbf{Age Group} & & & \\
        19–24                    & 2  & 0  & 2 \\
        25–29                    & 1  & 1  & 2 \\
        35–39                    & 4  & 3  & 7 \\
        40–44                    & 3  & 1  & 4 \\
        45+                      & 4  & 1  & 5 \\
        \hline
        \textbf{Gender} & & & \\
        Male                     & 11 & 4  & 15 \\
        Female                   & 3  & 2  & 5 \\
        \hline
    \end{tabular}
    \label{tab:demographic_preference}
\end{table}

\subsection{Impact of Technical Proficiency on Usability}

Participants with higher technical proficiency completed token-based setup tasks faster and with fewer errors, suggesting that technical complexity remains a critical barrier for less experienced users. As one participant noted, \textit{``I had to check whether my computer was 32-bit or 64-bit, which was confusing.''} (Participant 12), highlighting how seemingly simple technical requirements can hinder successful installation.

Future implementations of token-based systems should focus on simplifying the setup process and offering clearer guidance to support a wider range of users.

\subsection{Security vs. Usability Trade-off}

The security versus usability trade-off was clearly evident in participants' feedback. This observation aligns with previous findings that authentication mechanisms often suffer from usability challenges, leading to user dissatisfaction and avoidance~\cite{bonneau2012}. 

Remote e-signatures, while more usable and convenient, raised security concerns related to password protection. One participant expressed this by stating, \textit{``Remote e-signatures are easier, but I always worry about my password being stolen.''} (Participant 8). This finding aligns with earlier studies indicating that users often prioritize convenience over strict security measures~\cite{sasse1999}.

Conversely, token-based signatures provided a higher sense of security through physical device possession but posed usability challenges, especially during setup and portability. Another participant commented, \textit{``Even if it takes longer to set up, I trust a device I can hold in my hand.''} (Participant 6). This user frustration is consistent with findings that users often rationally reject security measures perceived as excessively burdensome~\cite{herley2009}.

\subsection{Implications for Adoption}

Organizations prioritizing user convenience and scalability should consider remote e-signature systems, particularly when dealing with diverse user groups. Meanwhile, sectors with stringent security requirements may prefer token-based solutions despite their usability drawbacks.

Hybrid models that integrate strong authentication mechanisms into remote systems could help bridge the usability-security gap identified in this study. As one participant summarized, \textit{``It would be perfect if I could have the ease of remote signatures with the security of a token.''} (Participant 15).

\subsection{Security Perception and Authentication}

Several participants expressed concerns about password-based access being insecure and potentially vulnerable to unauthorized use. This concern was particularly evident in cases where no secondary code—such as a one-time passcode (OTP)—was requested during login, leading to the perception that entering a password alone was sufficient for accessing and signing documents.

In reality, platforms like DocuSign employ multiple layers of authentication to ensure document security. One-time passcodes are often used during identity verification before users can access or sign sensitive documents. These OTPs are typically delivered via SMS or phone call and remain valid for a limited time, usually around 10 minutes. This mechanism ensures that even if a password is compromised, unauthorized access is still prevented unless the attacker also has access to the user's phone.

Moreover, DocuSign and similar platforms support alternative authentication methods beyond passwords, including biometric verification, identity document validation, and knowledge-based authentication (KBA), depending on the configuration set by the document sender. Therefore, the belief that password entry is the sole gateway to digital signing contributes to a skewed sense of vulnerability. Addressing this gap in user understanding—through clear user communication and platform feedback—can help align perceived and actual security, ultimately improving trust in remote signature technologies.

\section{Limitations and Future Work}

\subsection{Study Limitations}
This study had a limited sample size of 20 participants, which may not fully represent broader user demographics. Additionally, all users were required to test both methods sequentially, potentially introducing bias in their evaluations.

\subsection{Future Research Directions}
Future studies should expand the sample size and include participants from diverse technical backgrounds. Additionally, evaluating hybrid models—such as biometric authentication in remote e-signatures—could provide deeper insights into balancing security and usability.

\section{Conclusion}

This study provides empirical insights into the usability and security perceptions of token-based and remote electronic signatures by analyzing acquisition, installation, and usage processes in a controlled user study.

The results indicate that:

\begin{itemize}
    \item Remote e-signatures are perceived as significantly more usable than token-based e-signatures.
    \item Token-based e-signatures are perceived as significantly more secure than remote e-signatures.
    \item Although more participants preferred remote e-signatures for signing, this preference was not statistically significant.
\end{itemize}

Task performance analyses revealed that installation complexity was a key barrier for token-based e-signatures, affecting the overall user experience. In contrast, remote e-signatures required minimal setup and allowed participants to complete tasks more easily and intuitively.

Qualitative feedback further emphasized the usability-security trade-off, reflecting user preferences based on individual risk tolerance and situational needs.

This study extends the systematized framework established by Çağal and Bıçakcı~\cite{cagal} by providing empirical insights into how real users interact with token-based and remote electronic signature systems. While previous research identified potential usability barriers through expert-driven analyses, our findings offer concrete validation through user-centered experiments. The observed security-usability trade-off, user preferences, and task performance results not only confirm many of the theoretical challenges previously outlined but also reveal new nuances in user behavior and expectations.

Overall, this work complements prior system-level evaluations by emphasizing the importance of integrating usability engineering principles into the design of electronic signature systems. Future research should continue this trajectory by exploring hybrid models that balance security and usability, informed by both conceptual systematizations and empirical user studies.

\subsection{Proposed Solutions for Improving Usability}

To enhance both usability and security in e-signature systems, the following recommendations are proposed:

\begin{itemize}
    \item Implement hybrid authentication models, combining biometrics and multi-factor authentication.
    \item Simplify the installation processes for token-based systems and explore web-based token authentication options.
    \item Develop user-centric interfaces with guided onboarding and real-time support features.
    \item Promote the consistent use of One-Time Passwords (OTP) as an additional security layer, especially in remote signing scenarios. OTPs offer a lightweight and widely familiar method to enhance trust without compromising usability.
\end{itemize}

Future work should expand the participant pool to include a more diverse user base and evaluate hybrid e-signature models that combine the strengths of both remote and token-based methods. Additionally, upcoming regulations such as the eIDAS 2.0 proposal are expected to facilitate wider adoption of remote Qualified Electronic Signatures across Europe~\cite{eidass2}. Initiatives like the European Digital Identity Wallet Pilot aim to enhance cross-border digital identity verification and will likely impact the usability and adoption of remote e-signature systems~\cite{eudiwallet2022}. Recent reports indicate that the adoption of electronic identification solutions across Europe continues to grow, as reflected in the Digital Economy and Society Index (DESI) 2023~\cite{desi2023}.

Overall, this study underscores the need to design e-signature solutions that effectively balance security and usability to meet evolving user demands.

\end{document}